\documentclass[conference]{IEEEtran}
\IEEEoverridecommandlockouts
\usepackage{cite}
\usepackage{amsmath,amssymb,amsfonts}
\usepackage{algorithmic}
\usepackage{graphicx}
\usepackage{textcomp}
\usepackage{xcolor}
\usepackage{hyperref}
\usepackage{multirow}
\def\BibTeX{{\rm B\kern-.05em{\sc i\kern-.025em b}\kern-.08em
    T\kern-.1667em\lower.7ex\hbox{E}\kern-.125emX}}
\begin{document}

\title{Evaluating the Usability of LLMs in Threat Intelligence Enrichment\\
}

\author{\IEEEauthorblockN{1\textsuperscript{st} Sanchana Srikanth}
\IEEEauthorblockA{\textit{School of Computer Science} \\
\textit{University of Guelph}\\
Guelph, Canada \\
sanchana@uoguelph.ca}
\and
\IEEEauthorblockN{2\textsuperscript{nd} Mohammad Hasanuzzaman}
\IEEEauthorblockA{\textit{School of Computer Science} \\
\textit{University of Guelph}\\
Guelph, Canada \\
mhasanuz@uoguelph.ca}
\and
\IEEEauthorblockN{3\textsuperscript{rd} Farah Tasnur Meem}
\IEEEauthorblockA{\textit{School of Computer Science} \\
\textit{University of Guelph}\\
Guelph, Canada \\
ftasnurm@uoguelph.ca}

}

\maketitle
\begin{abstract}
Large Language Models (LLMs) have the potential to significantly enhance threat intelligence by automating the collection, preprocessing, and analysis of threat data. However, the usability of these tools is critical to ensure their effective adoption by security professionals. Despite the advanced capabilities of LLMs, concerns about their reliability, accuracy, and potential for generating inaccurate information persist. This study conducts a comprehensive usability evaluation of five LLMs—ChatGPT, Gemini, Cohere, Copilot, and Meta AI—focusing on their user interface design, error handling, learning curve, performance, and integration with existing tools in threat intelligence enrichment. Utilizing a heuristic walkthrough and a user study methodology, we identify key usability issues and offer actionable recommendations for improvement. Our findings aim to bridge the gap between LLM functionality and user experience, thereby promoting more efficient and accurate threat intelligence practices by ensuring these tools are user-friendly and reliable.   
\end{abstract}


\section{\textbf{Introduction}}
The evolving landscape of the digital world has led to an unprecedented growth in cyber-attacks, posing significant challenges for many organizations\cite{Alam_2024}. Cyber Threat Intelligence (CTI), which involves the collection, analysis, and dissemination of information about potential or current threats to an organization’s cyber systems, can provide insights to help organizations defend against these attacks\cite{Alam_2024,1}. 

Large Language Models (LLMs) have the potential to revolutionize CTI by enhancing the ability to process and analyze vast amounts of unstructured threat and attack data, allowing security analysts to utilize more intelligence sources than ever before\cite{Karabacak_Margetis_2023}\cite{Poireault_2024}. As sophisticated models, LLMs are frequently employed to provide a comprehensive understanding of textual data, aiding analysts in extracting valuable insights from a vast sea of information\cite{CTIcite,2}. However, despite their advanced capabilities, LLMs are prone to hallucinations and misunderstandings of text, especially in specific technical domains. These issues can lead to the generation of false or unreliable intelligence, which could be disastrous if used to address real cyber threats. This necessitates a careful consideration of using LLMs in CTI, as their limitations could undermine their effectiveness and reliability in this critical domain. 

As the scope of digital threat continues to evolve, so too must the tools we use to understand and counteract these threats. LLMs, developed initially for broad applications, are now being fine-tuned to meet the specialized needs of cybersecurity. The lack of proper benchmark tasks and datasets to evaluate LLM capabilities in CTI leaves their reliability and usefulness an open research question. Existing benchmarks provide general evaluations of LLMs but fail to capture the practical and applied aspects of specific CTI tasks. General benchmarks like GLUE\cite{Glue}, SuperGLUE\cite{Superglue}, and various domain-specific benchmarks provide datasets and frameworks for evaluating LLMs in terms of general language understanding or domain-specific capabilities\cite{Poireault_2024}. However, these benchmarks do not adequately address the practical aspects of cybersecurity. 

Given these challenges, it becomes imperative to conduct a thorough usability evaluation of LLMs in the CTI domain. This study focuses on five LLMs—ChatGPT, Gemini, Cohere, Copilot, and Meta AI—and aims to assess their capabilities in automating the collection, preprocessing, and analysis of threat data. By improving the usability of LLMs, security professionals can better leverage these advanced tools to defend against the ever-evolving cyber threats. 

\textit{The contributions of this paper are:}\begin{itemize}
\item Comprehensive Usability Evaluation: A thorough usability evaluation of five state-of-the-art LLMs—ChatGPT, Gemini, Cohere, Copilot, and Meta AI—specifically focusing on their capabilities in threat intelligence enrichment. 
\item Identification of Usability Issues: Heuristic walkthrough and user study to identify key usability issues that affect the effectiveness and efficiency of LLMs in the context of threat intelligence. 
\item Actionable Recommendations: Actionable recommendations for improving the usability of each evaluated LLM based on detailed analysis.
\item Benchmarking and Analysis: Bridge the gap between LLM functionality and user experience by providing insights into the strengths and weaknesses of LLMs in CTI tasks, thus promoting more efficient and accurate threat intelligence practices. 
\end{itemize}
\section{\textbf{Related work}}

The integration of Large Language Models (LLMs) into cyber threat intelligence (CTI) practices has the potential to revolutionize the field by enhancing the analysis of unstructured threat data \cite{3,4}. However, as discussed in “Ctibench: A benchmark for evaluating llms in cyber threat intelligence” by M. Tanvirul Alam, D. Bhushl, L. Nguyen, and N. Rastogi\cite{relatedwork-1}, LLMs are not without their limitations, including a tendency to hallucinate and misunderstand text, which can impact their reliability in generating accurate intelligence. To address this, the authors propose CTIBench, a suite of benchmark tasks and datasets, to evaluate LLMs specifically for CTI tasks, including vulnerability assessment and threat report attribution. This work aims to improve the usability and effectiveness of LLMs in CTI, ensuring more efficient and accurate threat intelligence practices. 

Building on this, “Evaluation of llm chatbots for osint-based cyberthreat awareness” by M. Hassanin and N. Moustafa\cite{relatedwork-2} explores the potential of LLMs in the form of chatbots for cybersecurity and CTI. The paper investigates the performance of various paid and open-source LLM-based chatbots, such as ChatGPT and Stanford Alpaca, in binary classification and named entity recognition (NER) tasks using a CTI dataset. The study highlights the flexibility and adaptability of these chatbots to CTI requirements and evaluates their performance and cost implications. While chatbots show promise, there is a need to critically assess their capabilities and limitations in generating accurate and reliable intelligence. 

In the broader context of cyber security, “A comprehensive overview of large language models (llms) for cyber defences: Opportunities and directions” by Hassanin, M., \& Moustafa\cite{relatedwork-3} emphasizes the expanding scope of cyber threats with the integration of technologies like the Internet of Things (IoT) and the need for advanced cyber defense strategies. The paper discusses the impact of LLMs on cyber defense, including their ability to process large volumes of data and generate human-like text. However, the survey primarily focuses on LLMs for threat intelligence and privacy preservation, among other categories, leaving a gap in evaluating their usability and performance in CTI-specific tasks. 

To address this gap, Actionable cyber threat intelligence using knowledge graphs and large language models” by R. Fieblinger, M. T. Alam, and N. Rastogi\cite{relatedwork-4} proposes SecurityBERT, a lightweight and privacy-preserving architecture for cyber threat detection in IoT networks. The model utilizes a novel encoding technique, Privacy-Preserving Fixed-Length Encoding (PPFLE), to efficiently represent network traffic data. SecurityBERT achieves impressive accuracy in multi-category classification of cyber threats, outperforming other ML algorithms. While this demonstrates the potential of LLMs in cyber threat detection, it is important to critically evaluate their performance in more complex and diverse CTI tasks, especially those involving unstructured text analysis. 

While these studies contribute to the understanding of LLMs in cyber security and CTI, there is a lack of comprehensive usability evaluation of LLMs specifically for threat intelligence enrichment \cite{5,6}. The current study aims to bridge this gap by conducting a heuristic walkthrough and user study to identify usability issues and provide actionable recommendations for improving the user experience and reliability of LLMs in CTI practices. Unlike previous work, our study focuses on the user interface design, error handling, learning curve, performance, and integration of LLMs with existing tools, ensuring a more practical and applicable assessment of their capabilities in real-world CTI scenarios. 

\section{\textbf{Methodology}}
In this section, we first justify our choice of Large Language Models (LLMs) and describe their interfaces. Next, we outline the study environment, including the data, set of tasks and criteria used to evaluate the LLMs. We also describe our approach to conducting the heuristic walkthroughs\cite{Heuristic} with the help of two phases- Task Oriented Evaluation and Free Form Exploration, and user studies.  

\subsection{\textbf{Large Language Models (LLMs) }}\label{AA}
We chose to evaluate five LLMs: ChatGPT, Gemini, Cohere, Copilot and Meta AI. This section justifies our choice of these LLMs and describes their interfaces, focusing on how they present information to users and their integration with threat intelligence workflows. 

We considered a wide range of LLMs from various providers, from a curated list of Large Language Models\cite{LLMs}. To narrow the selection of LLMs for our evaluation, we followed two main criteria:  

The first one was Availability and Accessibility: We only considered tools that we could access and run. This criterion limited our selection to both commercial and open-source tools available for evaluation. Secondly, Relevance to Threat Intelligence: On identifying tools that offer a broad spectrum of functionalities and integration options, we shortlisted them to evaluate their usage for threat intelligence enrichment. 

We selected LLMs known for their potential in handling cybersecurity-related tasks, ensuring they are suitable for threat intelligence applications. While definitive usage statistics for LLMs are not always publicly available, we can characterize the tools based on their popularity and adoption across various sectors. Tools like ChatGPT and Co-pilot benefit from OpenAI's extensive developer community and partnerships with major technology companies, indicating widespread use in both enterprise and academic settings. Cohere and Meta AI are favored for their enterprise-grade solutions in data analysis and machine learning applications, appealing to organizations seeking advanced AI capabilities. Tools with open-source components, such as Gemini, often gain traction through community contributions and integrations into popular software development workflows. While commercial tools like Cohere and Meta AI offer advanced features, open-source alternatives like ChatGPT and Gemini provide flexible solutions that appeal to a broad user base. 

\subsubsection{ChatGPT}
ChatGPT is a language model developed by OpenAI, based on the GPT architecture. It serves as a versatile tool for natural language processing tasks, including text generation, translation, and information retrieval. We utilized the publicly available API version of the chatbot for our evaluations. 

The main interaction pane allows users to input text queries and receive generated responses. The model's responses are displayed in real-time, with options for users to explore detailed outputs and adjust parameters such as response length and tone. Additionally, the interface includes a 'Settings' menu where users can customize language preferences and API integration settings\cite{7}. 

ChatGPT's usability features include intuitive error handling and feedback mechanisms. It provides error messages and suggestions when input queries are ambiguous or unsupported. The model's interface also supports contextual navigation, enabling users to trace back previous interactions and review detailed logs of past queries. Chat GPT can provide personalized responses by remembering user preferences and tailoring its responses accordingly. This feature can help create a more engaging and satisfying user experience, as users feel the system can understand and respond to their unique needs\cite{Chatgpt}. ChatGPT's customizability is another critical advantage. It can be fine-tuned to perform specific tasks or applications, such as customer service or language translation, by adjusting its training data and algorithms\cite{ChatGPTM}. Since ChatGPT is developed as a conversational agent, it allows a user to correct it when it makes a mistake. However, at times, ChatGPT does not accept the user’s correction and is over confident in its initial response\cite{ChatGPTM}. 
\begin{figure}[htbp]
  \centering
  \includegraphics[width=\columnwidth]{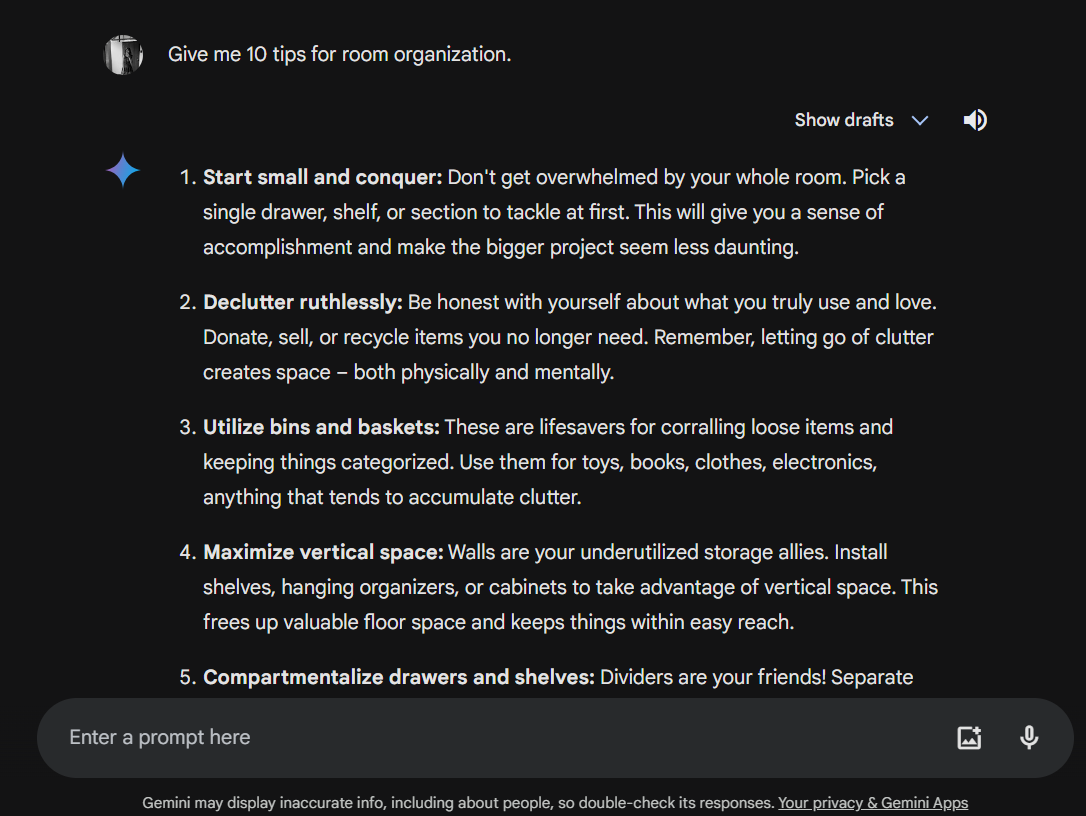}
  \caption{The User Interface of Gemini.}
  \label{fig:geminiui}
\end{figure}
\subsubsection{Gemini}
Gemini is the next generation of large language models (LLMs) developed by Google \cite{Gemini1}. It can understand text, images, videos, and audio. As a multimodal model, Gemini excels in complex tasks in various programming languages, including math, physics, and high-quality code generation, and it is currently available through the Gemini chatbot (formerly Google Bard)\cite{Gemini}. 
Figure \ref{fig:geminiui} shows the Gemini user interface. The main interface allows users to enter the query in the text area and send it to the chatbot to show its responses in real-time. Also, users can ask Gemini to create an image and generate the image on the same window \cite{Gemini}.  

The most powerful feature of the Gemini is its ability to understand and deal with various types of data such as text, images, audio, PDFs, and videos and generate more complete answers that fit the context\cite{Gemini3}. Furthermore, Gemini has significant potential for advancing educational technology and implementing practical applications beyond its theoretical framework \cite{Gemini4}. The Gemini excels across various domains, offering multimodal support for language learning, object recognition, diverse inputs, and real-time conversation, benefiting users with limited access to digital tools. 
\begin{figure}[htbp]
  \centering
  \includegraphics[width=\columnwidth]{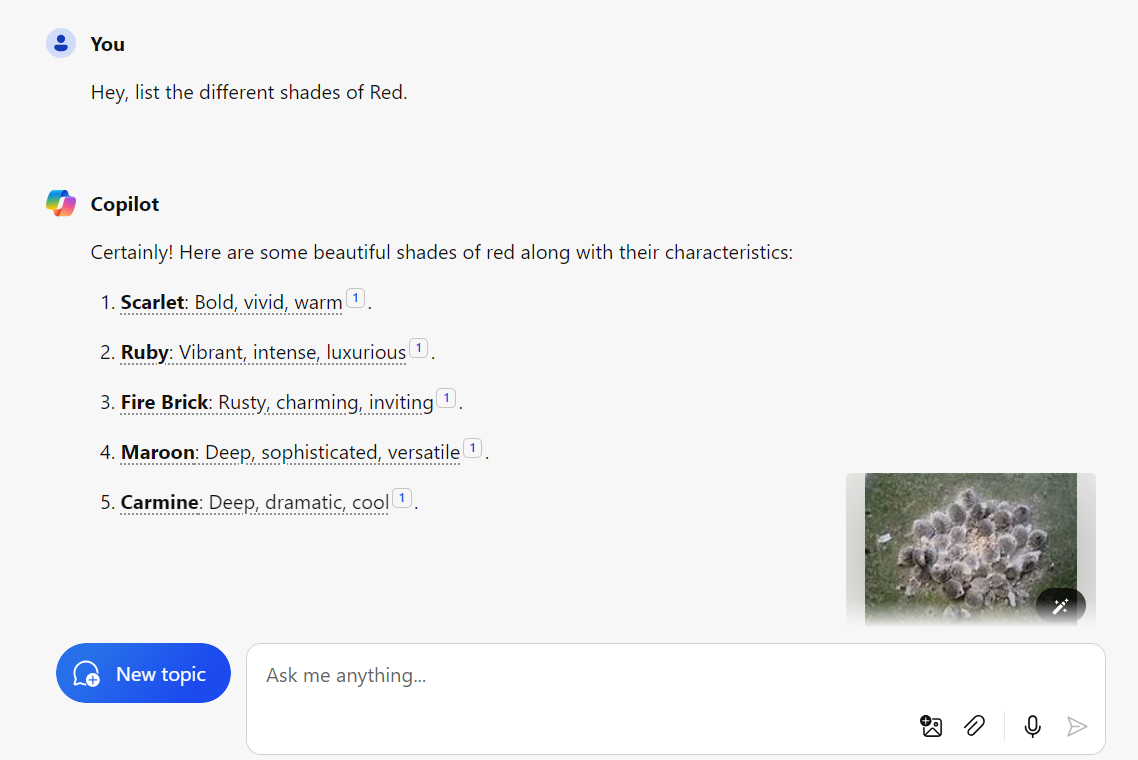}
  \caption{The User Interface of Copilot.}
  \label{fig:copilotui}
\end{figure}
\subsubsection{Cohere}
Cohere offers a variety of language models designed for different use cases. Some of their key models are Command, Command R, and Command R+. Command is the default instruction-following conversational model \cite{cohere}. 

Cohere is an Enterprise AI platform that empowers developers and enterprises to create applications using Large Language Models (LLMs). To get started, obtain an API key from the Cohere dashboard and install the Python SDK for Cohere. Verify the installation by running a simple script. For content generation, utilize the /chat endpoint to generate personalized content, adjusting parameters like model, prompt, and temperature for tailored responses. Additionally, you can fine-tune models by uploading a dataset to create custom models for generative, classification, reranking, or chat tasks. Cohere also offers semantic search capabilities by leveraging embeddings, which enable you to build search systems that outperform traditional keyword-based searches by incorporating context and user intent \cite{cohere}. 

\subsubsection{Copilot}
Copilot is a specialized language that is embedded into the Haskell functional programming environment. Its primary purpose is to rigorously monitor networked real-time systems \cite{copilot}. It works by creating monitors that continuously examine and validate software behaviors, which is vital for applications in ultra-critical systems like avionics and air-ground data lines. Copilot's declarative design assures that all expressions have no side effects, increasing clarity and maintaining referential transparency throughout its specifications. This method enables for exact monitoring of system variables and functions using streams, which are infinite sequences of values that conform to certain kinds.

Figure \ref{fig:copilotui} illustrates the Copilot interface, showcasing its integration into the Haskell programming environment. Users interact with Copilot through a structured development environment where they define streams of data and specify monitoring properties using Haskell's functional syntax. The interface facilitates the definition of temporal properties and the integration of monitoring components into existing systems seamlessly. It supports the scheduling of monitoring tasks and the visualization of monitored data streams in real-time, enhancing the system's fault detection capabilities and overall reliability. Copilot's use of Haskell as a foundation ensures scalability, making it suitable for complex, safety-critical applications where runtime verification is essential for system integrity and performance. 

\subsubsection{Meta AI}
Meta AI's Open Pretrained Transformer (OPT) is a cutting-edge language model developed by Meta AI. OPT has an astonishing 175 billion parameters learned on publicly available datasets, allowing it to excel at writing human-like language. OPT's principal applications include text production, handling simple math problems, completing reading comprehension questions, and simulating human-like interaction. OPT's characteristics make it a useful tool for a variety of natural language processing jobs\cite{Metaai}.

Through the interface, users can input text prompts and instantly receive generated responses. Users can modify reaction time and style, among other parameters, to meet their needs. To further enhance user experience, the UI has a settings section where users may adjust language preferences and other settings. Contextual navigation is supported by OPT's interface, giving users the ability to go back and edit past interactions, making for a smooth and simple user experience. 

OPT is a vital tool for activities ranging from customer service automation to content generation because of its strengths, which rest in its capacity to generate coherent and contextually relevant language based on minimal input. Furthermore, OPT's open-source design encourages innovation and advancements in AI-driven text production by enabling researchers and developers to play with and enhance the model\cite{Metaai}. 

\subsection{\textbf{Heuristic Walkthroughs}}
To identify usability issues in LLMs for threat intelligence enrichment, we employed a heuristic walkthrough methodology, combining cognitive walkthroughs\cite{cognitive,8} and heuristic evaluations\cite{HeuristicE}. This approach leverages the strengths of both techniques: cognitive walkthroughs simulate user tasks, while heuristic evaluations systematically assess the system against established heuristics\cite{Heuristic,9}. 

Participants in this method make two passes through the LLMs. The first pass uses ‘thought-provoking' questions and requires the evaluators to work through a set of prioritized tasks. The second pass requires evaluators to use a set of heuristics to find additional problems. The assumption here is that the task-based review will enhance the heuristic review. Heuristic walkthroughs have demonstrated efficacy in various other domains including electronic health records systems. Two authors conducted the heuristic walkthroughs for all five LLM tools (ChatGPT, Gemini, Cohere, CoPilot, Meta AI), drawing on their extensive experience in usage of advanced threat intelligence tools and conducting laboratory experiments in cybersecurity. A guide to our heuristic walkthrough is available in Appendix A. 

\subsubsection{\textbf{Phase 1: Task-Oriented Evaluation}}
Phase 1 of a heuristic walkthrough is like a cognitive walkthrough, where evaluators approach a system with a list of predefined tasks. In this phase, evaluators will familiarize themselves with the system and use the LLMs (ChatGPT, Gemini, Cohere, Copilot, Meta AI) to perform tasks that actual users will try to complete. The aim is to identify usability issues that might arise during regular use of these tools for analyzing the CTI data. To do so, we used the following set of tasks: 
\begin{itemize}
\item Upload the XML file containing CTI data into the LLM 

\item Direct the LLM to extract all threat intelligence references from the data and categorize them 

\item Identify key Attacker Activities and Patterns and summarize into a report using the LLM, with the help of the categorized data. Please ensure that the report summarizes all key attacker activities from the event data. 

\item Assess the Quality of the Enrichment.
\end{itemize}

Each task's success criteria were based on the completion of the task and the clarity and accuracy of the results provided by the LLMs. To aid us with critical thinking of each LLM, we adapted guiding questions inspired by Sears’ list of guiding questions\cite{Heuristic,10}. These questions ask evaluators to consider whether users will: know what to do next; recognize and use the appropriate commands to request enrichment; evaluate the LLM's responses accurately; and receive clear feedback on the enrichment process. During Phase 1, we recorded all of our findings, particularly our impressions of the tool, and any usability issues we encountered. 

\subsubsection{\textbf{Phase 2: Free-Form Evaluation}}
In Phase 2, evaluators freely explored the LLMs using predefined heuristics to identify usability issues. This phase aims to uncover comprehensive usability problems that might not be apparent during task-oriented evaluations. 

For this evaluation, we utilized Nielsen's 10 Usability Heuristics\cite{Nielsen}, adapted to suit the demands of analyzing CTI data with LLMs. Our evaluators meticulously examined each model's interface, functionalities, and user interactions. The evaluation focused on critical aspects such as the clarity of system status updates, consistency in interface design, error mitigation capabilities, and the adaptability of LLMs across varying user proficiency levels. Nielsen’s 10 Usability Heuristics\cite{Nielsen} were tailored particularly to the context of LLMs in threat intelligence for our study, to identify usability issues with the help of pre-defined heuristics:

\begin{itemize}
\item Visibility of System Status 

Ensure that users are informed about what the LLM is doing at any given time, such as processing a query or fetching data. 

\item Match Between System and the Real World 

Use language and concepts that are familiar to threat intelligence professionals, avoiding technical jargon that may not be widely understood. 

\item User Control and Freedom 

Provide users with the ability to easily undo or redo actions, and allow them to navigate freely through the system without being locked into specific workflows. 

\item Consistency and Standards 

Ensure that similar actions and terminology are consistent throughout the tool, following industry standards and best practices for threat intelligence platforms. 

\item Error Prevention 

Design the system to prevent errors from occurring in the first place, such as offering confirmation prompts before performing potentially destructive actions. 

\item Recognition Rather Than Recall 

Minimize the user's memory load by making information, actions, and options visible and easily accessible. Provide autocomplete and suggestion features where appropriate. 

\item Flexibility and Efficiency of Use 

Cater to both novice and experienced users by providing shortcuts and advanced options for experienced users while keeping the interface simple for beginners. 

\item Aesthetic and Minimalist Design 

Avoid clutter and provide a clean, focused interface that presents only the necessary information and actions to the user. 

\item Help Users Recognize, Diagnose, and Recover from Errors 

Provide clear and informative error messages, and offer suggestions for how to resolve the issues encountered. 

\item Help and Documentation 

Provide comprehensive help and documentation that is easily accessible. Include tutorials, FAQs, and context-sensitive help within the tool. 
\end{itemize}

Throughout this phase, qualitative data was meticulously collected to capture usability challenges, interface discrepancies, and user interaction complexities observed during engagements with the LLMs. This data underwent thematic analysis to categorize findings into themes and subthemes, providing a structured foundation for reporting in Section 4. By leveraging Nielsen's established heuristics, we ensured a comprehensive evaluation of how well LLMs facilitate and enhance threat intelligence enrichment tasks, offering actionable insights for optimizing their usability and efficacy in practical applications. 

\subsection{\textbf{User Study}}
In the second part of our evaluation, we conducted a user study to triangulate the observations made in the heuristic evaluation. We recruited participants with varying degrees of experience in cybersecurity and threat intelligence analysis, who we refer to as \textbf{P01–P10}. Participants provided their consent to participate in the study and answered preliminary questions about their background and familiarity with threat intelligence. Participants answered questions on a grade scale of 1 (novice) to 5 (expert) about their experience. Table \ref{participantsummary} reports on their responses. In summary, participants reported their familiarity with threat intelligence (median 4/5), automation tools (median 3/5), and use of LLMs (median 2/5). 

Participants were presented with two of the five LLMs and tasked with using them to enrich threat intelligence events, by directing the LLMs to extract key attacker activities from the provided data.  Participants used both LLMs for approximately 25 minutes each, thinking aloud while performing the tasks. We conducted semi-structured post-task interviews to gather detailed feedback on their experiences. Appendix D has the list of post-study questions we used as a guide to this interviews.

During the study, participants were allowed to skip events they were uncomfortable analyzing. This approach accounted for differences in familiarity with threat intelligence and LLM tools. It also simulated real-world scenarios where analysts might selectively focus on certain threats. All participants managed to find events they were comfortable with. In the post-task interviews, participants described their experiences with the tools, focusing on how the tools helped/did not help them understand and perform the tasks with the given data. Appendix C has the list of questions we used to guide this discussion.  

To reduce fatigue, we only asked participants to interact with two tools. A Latin-square design\cite{Latin} was applied to avoid learning effects between the two tools. 

\subsubsection{\textbf{Data Extraction from User Study}}
We captured screen recordings, audio, and questionnaire responses from each session. Two of the authors reviewed the recordings to identify usability issues encountered by the participants. This process yielded a total of 140 individual usability incidents. We filtered these to retain only the intersection of both authors’ reports, reducing the number of total usability incidents to 52. Next, two authors classified these incidents into distinct usability categories using an open card sort methodology\cite{opencardsort,11}, which ultimately indicated a high level of agreement between all the participants. The two raters then discussed and agreed on a final classification, which we present in Section 4.

\begin{table}[htbp]  
\centering     
\caption{Number of Data} 
\label{tab:Datatable}  
\begin{tabular}{|c|c|c|c|}
\hline                 
Year & Total & Report & Malware \\ 
\hline
2008 & 191 & 2 & 0 \\ 
2009 & 105 & 2 & 0 \\
2010 & 800 & 7 & 32 \\
2011 & 3,340 & 14 & 319 \\
2012 & 4,524 & 22 & 465 \\
2013 & 19,571 & 47 & 1,798 \\
2014 & 18,842 & 100 & 1,116 \\
2015 & 17,258 & 78 & 1,554 \\
2016 & 543,703 & 79 & 1,974 \\
2017 & 15,660 & 72 & 1,017 \\
2018 & 13,210 & 125 & 1,300 \\
2019 (-Jun) & 5606 & 64 & 628 \\
Total & 642,810 & 612 & 10,203 \\
\hline
\end{tabular}
\end{table}

\subsubsection{\textbf{Dataset Description}}
\begin{figure}[htbp]
  \centering
  \includegraphics[width=\columnwidth]{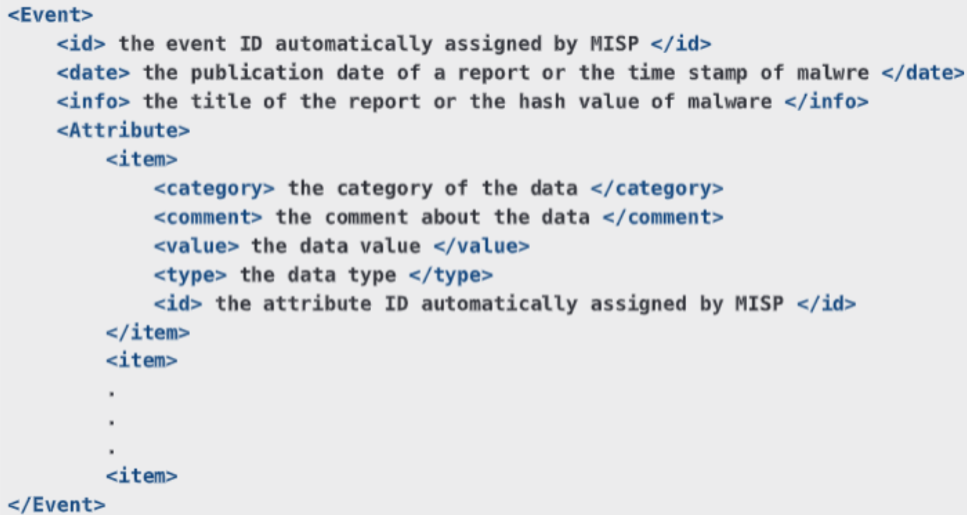}
  \caption{Data Schema of an Event}
  \label{fig:event}
\end{figure}

Evaluations were conducted using the ‘Cyber Threat Intelligent (CTI) dataset generated from public security reports and malware repositories’ \cite{Dataset}. This dataset was contributed by the authors particularly to aid in CTI analysis. The dataset is composed of several sets of events, and the data schema of an event is presented in Figure \ref{fig:event}. The dataset is stored in a structured format (XML) and includes approximately 640,000 records from 612 security reports published from January 2008 to June 2019. The amount of data, the report, and the malware events are shown in Table \ref{tab:Datatable}. Several data types are contained in this dataset such as URL, host, IP address, e-mail account, hashes (MD5, SHA1, and SHA256), common vulnerabilities and exposures (CVE), registry, file names ending with specific extensions, and the program database (PDB) path \cite{Dataset}.

The dataset was split and prepared as follows:
\begin{itemize}
    \item \textit{Training Set:} 70\% of the data was used for training the LLMs. This subset was used to train the LLMs to recognize patterns and learn from the threat intelligence data.
    \item \textit{Validation Set:} 15\% of the data was used for validation during the training phase.
    \item \textit{Testing Set:} The remaining 15\% was used for testing the performance of the LLMs.
\end{itemize}

Utilizing this dataset, we evaluate the ability of our chosen state-of-the-art LLMs to preprocess and organize the collected data, analyse them, and extract key attacker activities.   

\section{\textbf{Results}}
Through our heuristic evaluations and user study, we identified 70 and 52 usability issues, respectively. These issues highlight potential areas for improvement rather than reflecting the overall quality of the evaluated LLM tools. All identified issues from our evaluations and user study have been grouped by themes and sub-themes and produced in this section.

In each section, we describe the usability issues related to each theme, explain their impact on users, and suggest how these insights could be utilized by tool designers and researchers to enhance the usability of Large LLMs in threat intelligence enrichment. The number of usability issues identified during the heuristic evaluation phase is reported next to each theme title and this number is solely used to characterize our findings only. 

To further organize the results, we have bolded short titles describing subthemes of issues within each theme. Next to each subtheme title are the tools, in (brackets), that issues in that subtheme apply to. For instance, “Compatibility Issues (Copilot, Cohere, Gemini, MetaAI)” denotes that Compatibility issues apply to Copilot, Cohere, Gemini and MetaAI, but not ChatGPT. Table \ref{tab:themes} provides an overview of the themes and subthemes. 

\begin{table}[ht]
\centering
\caption{Usability issues grouped by theme} 
\label{tab:themes} 
\begin{tabular}{cc}
\hline
\textbf{Theme} & \textbf{Subtheme} \\ \hline
\multirow{2}{*}{A. Integration Challenges} & Compatibility Issues \\ 
                            & External Data Integration \\ \hline
B. Quality/Relevance of Output \\ \hline
\multirow{2}{*}{C. User Experience} & Visual Appeal \\  
                            & Error Message Displays \\ \hline
D. Operational Transparency \\ \hline
\multirow{2}{*}{E. Usability in Real-Time Operations} & Response Time \\ 
 & Memory Load \\
\end{tabular}
\end{table}

\subsection{\textbf{Integration Challenges (18)}}
Integrating LLMs with current threat intelligence systems presents substantial challenges that impact their usability in security environments. This theme examines the difficulties in merging these models with various industry-standard tools and data formats essential for cybersecurity. Key issues include the models' limitations in handling different file types and their inability to directly interact with external databases. These integration challenges significantly affect the efficiency, accuracy, and response speed of security analysts working with real-time data. Here, we focus on specific integration barriers, such as compatibility with diverse data formats and external data access, highlighting the need for LLMs to improve their data processing and connectivity features. 

\subsubsection{\textbf{Compatibility Issues (Copilot, Cohere, Gemini, MetaAI)}} The provided dataset in XML format could not be directly uploaded into Copilot and Cohere, necessitating manual data entry, while Gemini and MetaAI lacked file upload capabilities entirely. Gemini's ability to process image files did not extend to other file types, underscoring a significant flexibility issue. This limitation is critical as threat intelligence often requires the analysis of diverse data formats in real-time, restricting the usability of these tools in dynamic environments. The inability to handle various file formats natively means that users must often spend additional time on data preparation and conversion, which can delay the analysis process significantly. This lack of compatibility with common file types could lead to increased operational overhead and potential data integrity issues as information is manually transferred between formats. Moreover, the manual entry requirement increases the risk of human error, further compromising the reliability of the threat intelligence gathered. Such constraints underscore the need for broader data handling capabilities within these LLM platforms to accommodate the varied needs of threat intelligence analysis. 

\textit{User Study} : Similar issues were highlighted by the participants in the User study. Participants (P02, P03, P07, P08) expressed frustration with Cohere's inability to directly process the provided dataset. Similarly, Participants (P03, P04, P08) had issues with the restriction of Copilot not being able to take in the XML format dataset. Also, a Participant (P05) had remarked that Meta AI was not able to take files of any kind, and another(P07) stated that Gemini could take only images. 

\subsubsection{\textbf{External Data Integration (ChatGPT, Copilot, Cohere, Gemini, MetaAI)}} 
\begin{figure}[htbp]
  \centering
  \includegraphics[width=\columnwidth]{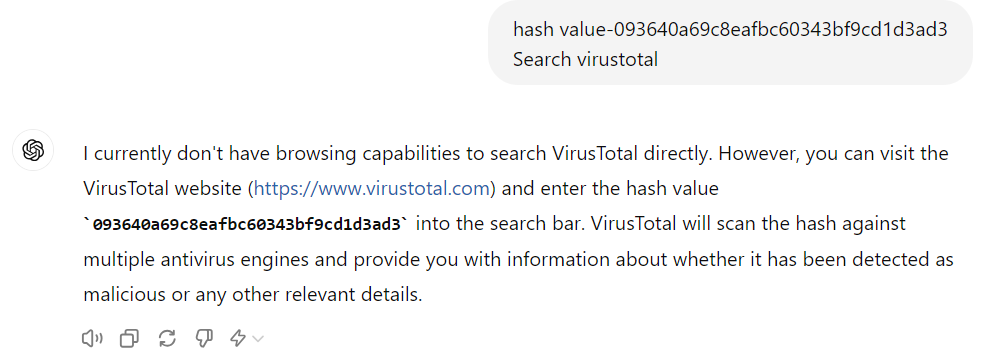}
  \caption{\textbf{External Data Integration} issue in ChatGPT}
  \label{fig:chatgptlimi}
\end{figure}

When tasked with verifying a hash value from an event in the dataset against VirusTotal, an external malware scanning service, all LLMs demonstrated a critical limitation: none could access external databases directly. ChatGPT's response typified this restriction, stating, “I currently don’t have browsing capabilities to search VirusTotal directly,” and merely offered steps for manual lookup. Figure \ref{fig:chatgptlimi} depicts this instance, where ChatGPT fails to retrieve data from an external source. This issue was consistent across other LLMs like Copilot, Cohere, Gemini, and MetaAI, all of which similarly lacked the capability to pull data from specific external resources such as VirusTotal. This limitation is particularly impactful in threat intelligence work, where the ability to quickly verify data against up-to-date external databases can significantly enhance the accuracy and reliability of threat assessments. The LLMs' inability to interface directly with external sources necessitates additional steps for analysts, potentially slowing down response times and affecting the efficiency of security operations. Additionally, this disconnect diminishes the potential for real-time intelligence gathering, placing a heavier workload on analysts to manually bridge the data integration gap.

\textit{User Study} : The issue of external data integration was not specifically highlighted by participants, possibly due to the nature of the prompts provided, which did not explicitly require external data access for task completion. 

\textbf{Discussion} : The issues identified in the integration and operational capabilities of these LLMs highlight a significant gap in their applicability to real-world threat intelligence tasks. Enhancing these models to handle a variety of data formats directly and to interact with external databases like VirusTotal could substantially improve their utility. To address these challenges, developers could focus on expanding the data handling capabilities of LLMs to include a wider range of file types and formats, thereby minimizing the need for manual data preprocessing. Furthermore, equipping LLMs with the ability to directly access and retrieve data from external sources would streamline workflows and increase the speed and accuracy of threat detection and analysis. By implementing these improvements, LLMs could better aid in threat intelligence enrichment.  

\subsection{\textbf{Quality/Relevance of Output (11) (MetaAI, Copilot, Cohere) }}
\begin{figure}[htbp]
  \centering
  \includegraphics[width=\columnwidth]{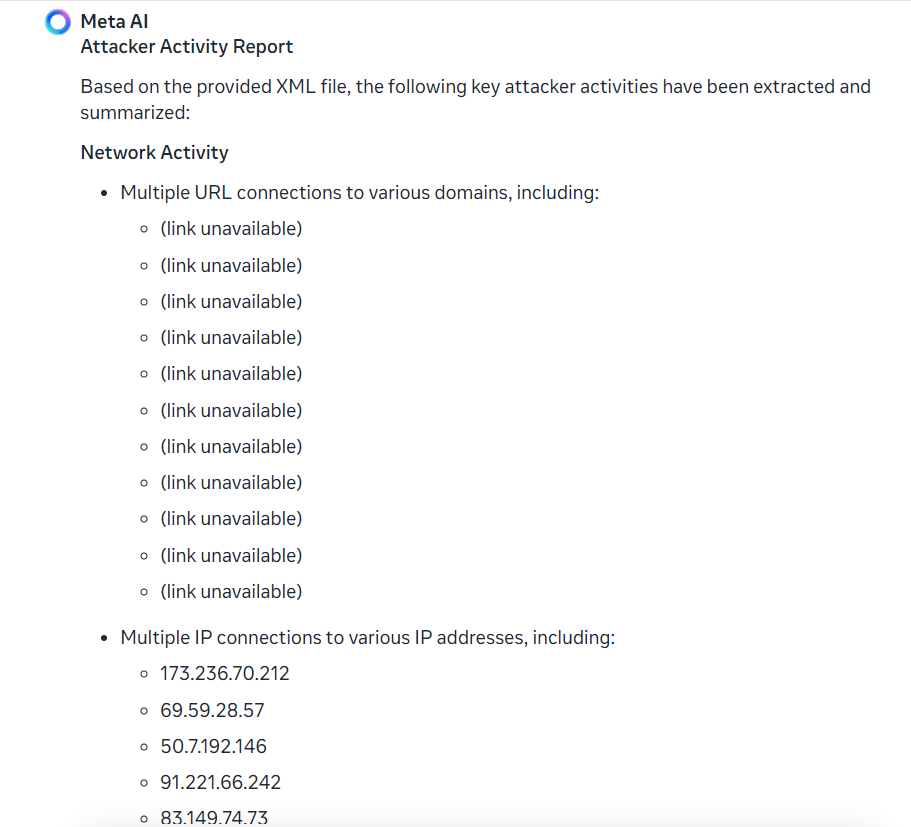}
  \caption{MetaAI's issue of \textbf{Quality/Relevance of Output}}
  \label{fig:metaaiqual}
\end{figure}

\begin{figure}[htbp]
  \centering
  \includegraphics[width=\columnwidth]{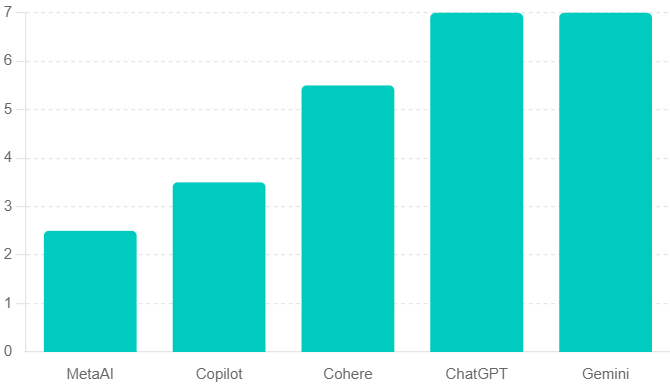}
  \caption{Performance of LLMs based on \textbf{Accuracy of Threat Data Categorisation}}
  \label{fig:outputqual}
\end{figure}

When tasked with categorizing the dataset and extracting key attacker activities, the output quality from MetaAI, Copilot, and Cohere varied significantly. MetaAI, for instance, failed to provide comprehensive reports and omitted links with a “link unavailable” error, falling short of expectations to classify URLs, file names, and IP addresses alongside a summary of key activities. Figure \ref{fig:metaaiqual} depicts this Quality of Output issue with MetaAI. Similarly, Copilot's output was underwhelming as it merely categorized the data without identifying or summarizing any attacker activities. In contrast, Cohere managed to summarize the key attacker activities but did not categorize the provided data as required. This inconsistency in performance highlights a crucial need for alignment in LLM capabilities with the specific demands of threat intelligence analysis. The varying degrees of output completeness and accuracy observed across these platforms while performing the tasks are depicted in Figure \ref{fig:outputqual}.

\textit{User study} : The shortcomings of these LLMs were consistently noted in the user study. Participants (P04, P09, P10) using MetaAI reported dissatisfaction with the incomplete and unsatisfactory nature of the reports. Participant (P02), who tested Copilot, pointed out the absence of any summarized findings, highlighting a gap in meeting the task requirements. 

\textbf{Discussion} : The observed inconsistencies in the outputs from MetaAI, Copilot, and Cohere highlight a significant challenge in the deployment of LLMs for threat intelligence tasks. The quality and relevance of the outputs are crucial for effective decision-making in cybersecurity contexts. MetaAI's incomplete reports and the missing link data suggest a need for improvements in data linkage and report generation mechanisms. Copilot's failure to provide summarized attacker activities points to a gap in its analytical processing, which could be addressed by enhancing its data interpretation algorithms. Cohere, while able to summarize key activities, still lacked in data categorization, indicating a need for a more robust data parsing and classification framework. Addressing these issues would not only improve the utility of each LLM but also enhance the overall reliability of LLMs in supporting cybersecurity professionals in real-time threat detection and analysis. 

\subsection{\textbf{User Experience (16) }}

\begin{figure}[htbp]
  \centering
  \includegraphics[width=\columnwidth]{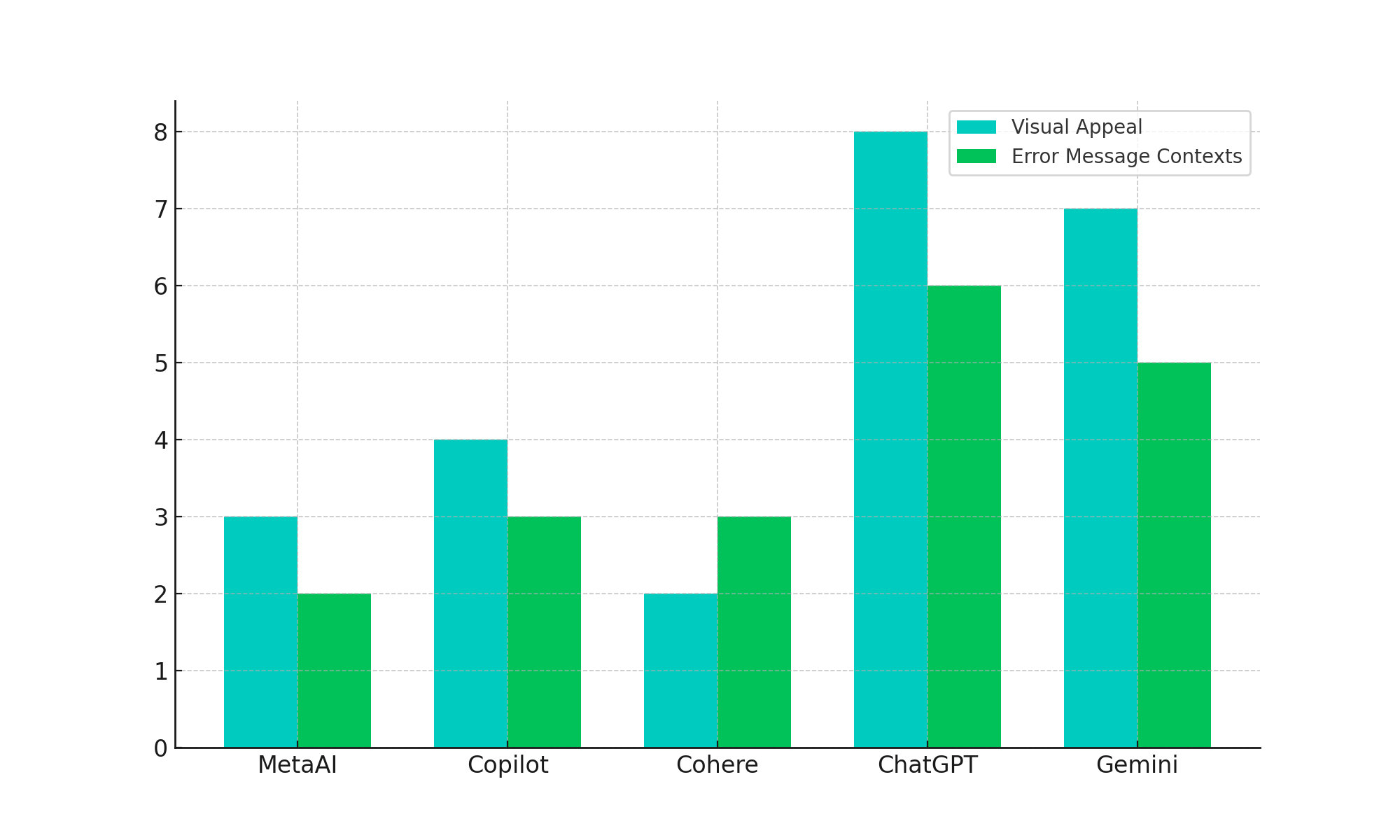}
  \caption{Comparison of LLMs based on \textbf{Visual Appeal and Context Of Error Message Displays}}
  \label{fig:userexpgraph}
\end{figure}

This section examines the aesthetic and functional aspects of user interfaces in the LLMs. An intuitive and visually appealing interface is crucial for facilitating effective user interaction and ensuring efficient task execution. Well-designed user interfaces can significantly enhance user satisfaction, reduce cognitive load, and improve overall productivity in threat intelligence operations. Figure \ref{fig:userexpgraph} shows how well each LLM was visually appealing, and compares their context of error messages displayed.

\subsubsection{\textbf{Visual Appeal (Cohere, Copilot)}} During the initial phase of our evaluation, Cohere's user interface significantly lacked visual appeal. The design elements employed did not allow an intuitive or engaging user experience, which is vital for ensuring user satisfaction and efficiency. Specifically, the text displayed on the dashboard was barely visible, appearing faded, which posed challenges to readability and interaction. Furthermore, the font style used was not only aesthetically unpleasing but also functionally impractical, complicating the task of quickly navigating through the interface. This type of visual presentation could potentially lead to increased user fatigue and decreased productivity. Moreover, the overall layout lacked the modernity and clarity expected in contemporary software tools, which could detract from user confidence in the tool's capabilities. Addressing these design flaws is crucial for improving user engagement and optimizing the operational effectiveness of the platform. 

\textit{User Study} : Participants (P02, P03) described Cohere’s design as ‘unappealing,’ highlighting issues with both the color scheme and the legibility of text. A participant (P09) noted similar shortcomings with Copilot, criticizing its monotonous interface and lack of dynamic visual elements that could otherwise enhance user interaction and make the analytical tasks more user-friendly. 

\begin{figure}[htbp]
  \centering
  \includegraphics[width=\columnwidth]{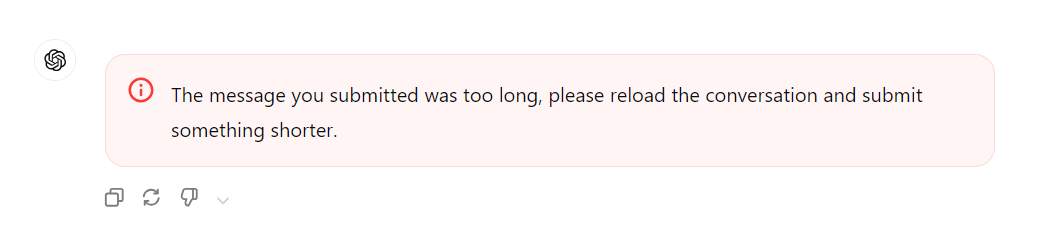}
  \caption{Insufficient context of \textbf{Error Message} displayed in ChatGPT}
  \label{fig:chatgpterror}
\end{figure}

\subsubsection{\textbf{Error Message Displays (MetaAI, ChatGPT, Cohere)}} 
\begin{figure}[htbp]
  \centering
  \includegraphics[width=\columnwidth]{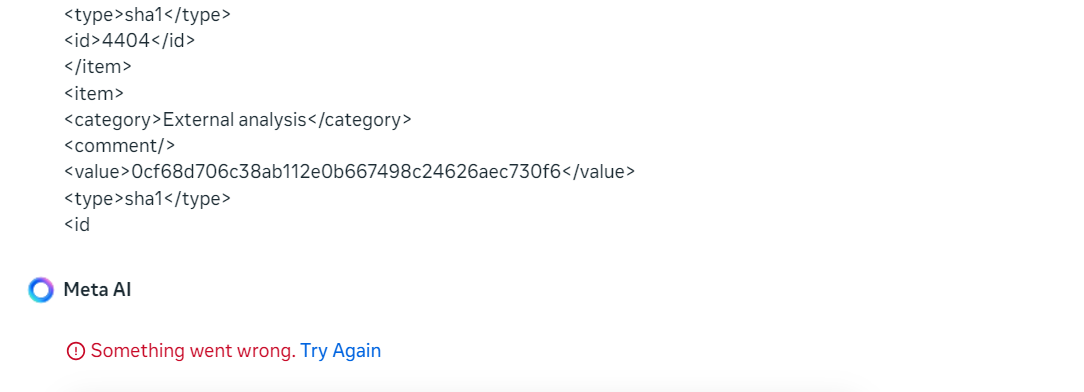}
  \caption{Insufficient context of \textbf{Error Message} displayed in MetaAI}
  \label{fig:metaaierror}
\end{figure}

MetaAI displayed a generic error message saying "something went wrong" when overwhelmed with large data inputs, which did not effectively communicate the nature of the issue to the user. A more descriptive error message, indicating specifics such as "input exceeds character limit," would have been more helpful for users to understand and rectify the input size accordingly. Similarly, while ChatGPT did inform the user saying, “The message you submitted was too long, please reload the conversation and submit something shorter,” it failed to provide guidelines on the acceptable input size, leaving users to guess the correct length through trial and error. This lack of detailed feedback can lead to increased user frustration and reduced efficiency as they may spend considerable time adjusting their inputs without clear direction. Figure \ref{fig:chatgpterror} and \ref{fig:metaaierror} illustrates this issue in ChatGPT and MetaAI respectively. Inconsistent error messaging across different LLM platforms complicates the user's ability to seamlessly interact with these tools, further hindering productivity in high-stakes environments. 

\textit{User Study} : One of the participants(P04) described the error message displayed by MetaAI as ‘blunt’ without proper context, which made troubleshooting more challenging. Another participant (P02) echoed this sentiment with Cohere, pointing out that it was ‘irritating’ to determine the acceptable input size through repetitive manual adjustments. 

\textbf{\textbf{Discussion}} : The identified usability issues with the LLMs' interfaces highlight the necessity for enhanced visual design and clearer error communication to improve user interaction and satisfaction. By prioritizing user-centered design principles and more descriptive, actionable error messages, developers can significantly enhance the operational effectiveness and user experience of these tools. Addressing these elements will not only reduce user fatigue and frustration but also streamline the workflow, making these systems more efficient and intuitive for threat intelligence analysts.  

\subsection{\textbf{Operational Transparency (8) (Copilot, Cohere, MetaAI) }}

During the heuristic phase-1 evaluation, when data was manually inputted into Copilot to extract key attacker activities from the provided CTI dataset, the system lacked explicit notifications about its processing status. This absence of feedback, such as a simple "generating response" indicator, left users uncertain about whether the system was processing their request or had stalled. Similar transparency issues were observed with Cohere and MetaAI, where there was a noticeable lack of updates on the actions being performed by the systems. This lack of communication is particularly problematic in real-time threat analysis contexts, where analysts require timely information to make informed decisions. In scenarios where timely and accurate threat intelligence is critical, these delays and uncertainties could hinder effective response strategies. Furthermore, the inability to ascertain system activity may lead to unnecessary repetitions of queries, compounding user frustration and system load\cite{12,13}. 

\textit{User Study} : Participants (P04, P07, P08, P09) found similar issues and expressed frustration with the systems' lack of responsiveness, with comments such as "not clear" and "unknown wait time" regarding the response times of Copilot and Cohere, respectively. Another Participant (P10) found transparency issues with MetaAI, stating that it was being “unclear in its actions.” 

\textbf{Discussion} : The challenges identified in operational transparency highlight a crucial gap in the design of these LLM systems, particularly in their failure to provide real-time feedback during data processing. To enhance the usability and effectiveness of LLMs in threat intelligence contexts, it is essential to incorporate clear, continuous status updates that keep users informed of the system's activity. This can prevent the redundancy of actions and reduce frustration by setting accurate expectations for response times. Implementing features such as progress bars, status indicators, or simple animations can provide visual cues that reassure users that their requests are being processed. Moreover, offering estimated times for task completions could further align the system's operations with the fast-paced requirements of threat intelligence tasks, thereby optimizing the workflow and enhancing user satisfaction and trust in the system's capabilities. 

\subsection{\textbf{Usability in Real-time Operations (17) }}

\begin{figure}[htbp]
  \centering
  \includegraphics[width=\columnwidth]{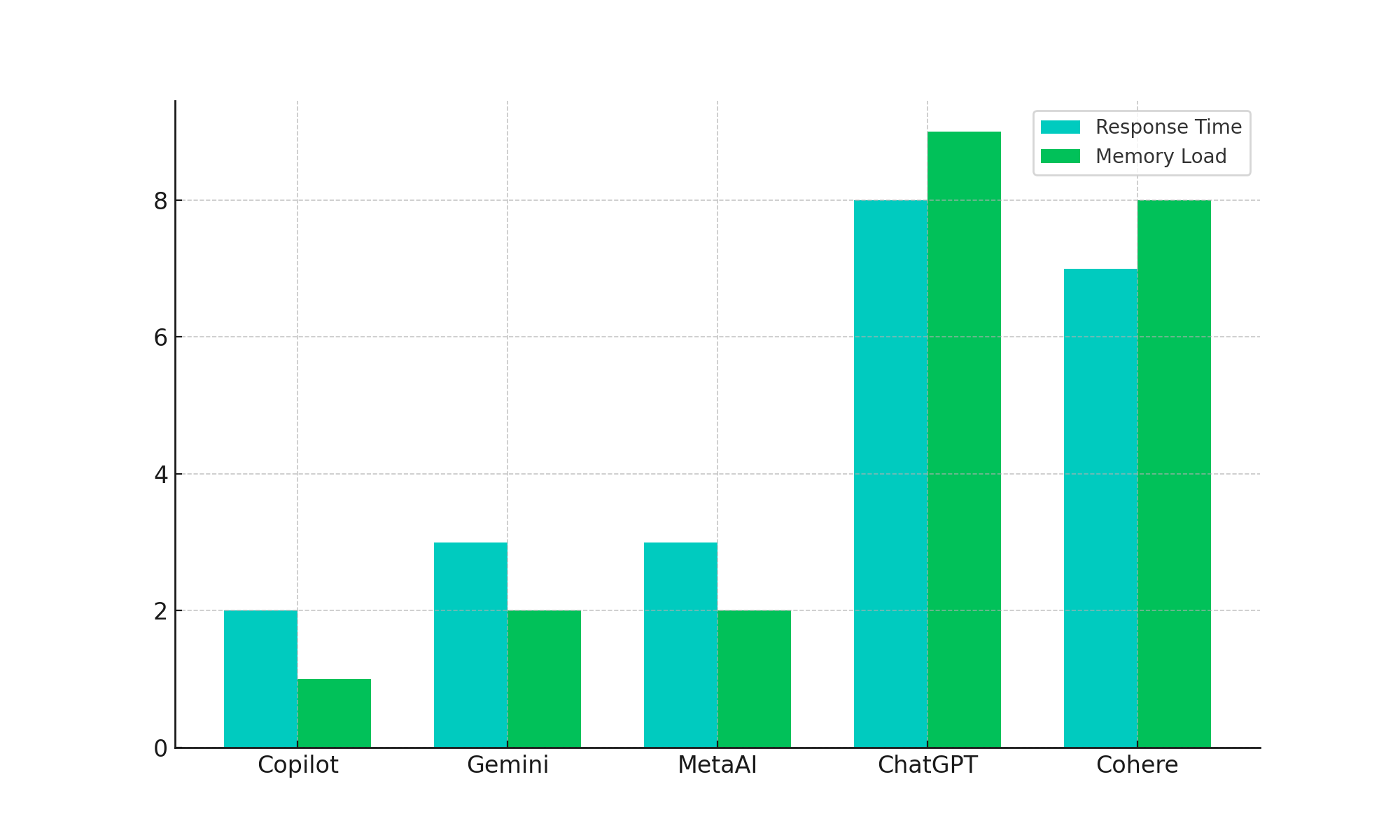}
  \caption{Performance of LLMs based on \textbf{Response Time and Memory Load}}
  \label{fig:perfgraph}
\end{figure}

This theme explores the practical usability of LLMs during real-time threat intelligence operations, focusing on how these systems manage response times and memory capabilities under the pressures of ongoing tasks. Efficient real-time operations require LLMs to provide quick responses and maintain continuity across sessions, critical for the fast-paced demands of threat analysis. The ability of LLMs to meet these requirements directly influences their effectiveness in supporting security analysts in active threat environments. Figure \ref{fig:perfgraph} is a visual representation of how each LLM performed in terms of response time and memory capabilities under the pressures of real-time threat intelligence operations- note that higher scores indicate better performance, with fewer issues.

\subsubsection{\textbf{Response Time (Copilot, Gemini, MetaAI)}} During the task-oriented evaluation, we observed that Copilot's response times to process and extract key attacker activities from the CTI dataset were considerably longer than anticipated. This issue was not isolated to Copilot; similar delays were experienced with Gemini and MetaAI. These extended response times are particularly concerning while using for threat intelligence analysis, where speed is often critical to the effectiveness of security responses. Comparatively, another LLM demonstrated significantly faster processing, highlighting a performance discrepancy that could impact user preference and operational efficiency. Prolonged delays in obtaining critical data can hinder timely decision-making in fast-paced environments, potentially compromising the effectiveness of threat mitigation strategies. 

\textit{User Study} : P01 faced this issue while using Gemini, stating the LLMs response time as ‘time-consuming’. Also, participants (P03, P08) found issues with the response time of Copilot. 

\subsubsection{\textbf{Memory Load (Copilot)}} In our evaluation of Copilot, we encountered significant limitations with its memory capabilities. It lacked the essential feature of recalling previous interactions, which became evident when we attempted to revisit earlier discussions to extract and verify information. This deficiency was particularly pronounced when we switched to a new conversation within the same session—previous conversations were not preserved. Furthermore, upon logging out or closing the browser tab, all conversational history was lost, posing a severe impediment to workflow efficiency. Such limitations are critically disadvantageous in the context of threat intelligence analysis, where the ability to recall and build upon past data is crucial. The necessity for users to manually retain and re-input data can lead to increased cognitive burden and potential inaccuracies, undermining the tool's overall effectiveness in dynamic security environments. Copilot’s inability to support continuous memory in interactive sessions significantly hampers its utility as a reliable tool for comprehensive CTI data enrichment tasks, where seamless access to historical data and insights is essential for informed decision-making. 

\textit{User study} : All participants (P03, P04, P08, P09) who worked with Copilot reported difficulties due to the system's inability to remember past interactions. 

\textbf{Discussion} : The challenges observed in the response times and memory retention of Copilot, Gemini, and MetaAI highlight significant areas for improvement to meet the demands of real-time threat intelligence analysis. Enhancing response efficiency and ensuring seamless session memory are critical for these tools to be truly effective in operational settings. Faster processing speeds would reduce downtime and enable analysts to act swiftly on intelligence insights. Moreover, integrating persistent memory capabilities would allow continuous analytical processes, reducing the cognitive load on users and preventing data loss during critical analyses.

\section{\textbf{Design Guidelines}}
To make our findings actionable and enhance the usability of LLMs in threat intelligence enrichment, we offer the following design guidelines based on the usability issues identified through our heuristic evaluations and user study.  
\begin{itemize}
\item \textbf{Support Diverse Data Formats and Integration.} LLMs should accommodate a wide array of data formats without requiring manual data preprocessing. This includes developing capabilities to automatically recognize and process different file types and integrating directly with external databases and services to streamline data retrieval and analysis. 

\item \textbf{Ensure Comprehensive and Actionable Outputs.} Enhance the LLMs’ ability to generate detailed, actionable reports that go beyond basic data categorization. Outputs should include comprehensive analysis of key attacker activities and provide actionable insights, directly linking to resources and supporting decision-making in threat intelligence. 

\item \textbf{Improve User Interface Design.} Optimize user interfaces to be more intuitive and visually appealing. This involves employing readable fonts, effective color schemes, and logical layout designs that reduce cognitive load, thereby increasing user satisfaction and operational efficiency. 

\item \textbf{Enhance Operational Transparency.} Implement real-time feedback mechanisms such as progress indicators and detailed status messages to keep users informed about what the system is doing. This will help set correct expectations for response times and reduce user anxiety during data processing. 

\item \textbf{Optimize Response Times for Real-time Operations.} Prioritize improvements in processing speed to ensure that LLMs can deliver prompt responses, which are critical in high-stakes environments like threat intelligence analysis. Quick feedback and rapid data processing are essential for timely decision-making. 
\end{itemize}
\section{\textbf{Limitations}}
Firstly, our evaluation was limited to five specific LLMs—ChatGPT, Gemini, Cohere, Copilot, and Meta AI. While these models are among the leaders in the field and offer a broad overview of current capabilities, they may not represent all LLMs, especially those specialized or less-known models that could be used in CTI tasks. To enhance the generalizability of our findings, we ensured that the selected models encompass a variety of functionalities and are representative of both commercial and open-source tools\cite{LLMs}. 

Another potential limitation is that the complexity and diversity of CTI tasks may have influenced the issues identified. The CTI domain involves various types of data and operational demands, and our scenarios, although designed to be representative, might not cover all potential real-world applications. This could affect the comprehensiveness of the usability issues we have reported. 

Our methodological approach also introduces certain constraints. The heuristic evaluation and user study were conducted within structured environments and specific tasks. This controlled setup might not completely capture the dynamic and often unpredictable nature of real-world CTI operations. Furthermore, the feedback and insights were gathered from a limited number of participants in the user study, which may not fully reflect the diversity of experiences and opinions in the broader cybersecurity community. 

The study’s reliance on manual data input, due to the LLMs' limitations in handling various file formats directly, presents another limitation. This might have influenced the evaluators' perception of usability and efficiency, potentially skewing the results toward identifying more issues related to data handling than might be present with better file support. 

While it would be beneficial to conduct a long-term study over several months within a real-world industry setting to observe the extended use and integration of LLMs in CTI operations, our current study's methodology of using heuristic evaluations already addresses this limitation to some extent. This approach allowed our evaluators to spend substantial time exploring the usability and integration of LLMs deeply. To mitigate potential external threats to validity and enhance the reliability of our findings, we carefully selected participants who have significant experience in the cybersecurity industry and are familiar with the demands and nuances of threat intelligence analysis. 

\section{\textbf{Conclusion}}
This paper addresses the critical need for comprehensive usability evaluations of LLMs in Threat Intelligence enrichment. Through a series of heuristic evaluations and a detailed user study, this research assessed the usability of five advanced LLMs: ChatGPT, Gemini, Cohere, Copilot, and Meta AI, specifically focusing on their capabilities to enhance threat intelligence processes. 

To promote better usability and to aid in the development of more effective LLMs for CTI, we have detailed a set of design guidelines based on our evaluations. These guidelines aim to help developers enhance LLM capabilities in handling diverse data formats, improving response times, and ensuring meaningful interaction with users. We hope that our research will not only assist practitioners in refining the usability of LLMs within the CTI field but also inspire further studies on the practical application and evaluation of LLMs in various domains of cybersecurity. 

\clearpage 
\bibliography{Ref}
\clearpage 
\bibliographystyle{IEEEtran}
\section*{Appendix}
\subsection*{\textbf{A Heuristic Walkthrough Guide}}
\textbf{Pass 1}

\textbf{Prioritized List of Tasks} 

1. Upload the XML file containing CTI data into the LLM 

2. Direct the LLM to extract all threat intelligence references from the data and categorize them 

3. Identify Key Attacker Activities and Patterns and summarize into a report using the LLM, with the help of the categorized data. Please ensure that the report summarizes all key attacker activities from the event data. 

4. Assess the Quality of the Enrichment.

Repeat these tasks with different events and different LLM tools until you feel satisfied with your assessments, to compare their performance and usability. Use the questions below to guide your evaluation. Record any usability problems you encounter during this phase.

\textbf{Guiding Questions}: 

1. Will users know how to select and load an event from the dataset into the LLM tool? Is the process of inputting the event data into the tool straightforward and intuitive? 

2. Will users recognize the appropriate commands or inputs to request an enrichment strategy from the LLM? Are the tool’s responses clearly presented, and do they include actionable insights? 

3. Will users be able to easily assess the quality and relevance of the LLM's responses? Does the tool provide sufficient feedback to confirm that the task is being performed correctly? 

4. Will users receive clear indications of progress as they interact with the tool? Is the feedback provided by the tool adequate for users to understand the effectiveness of the enrichment strategy? 

\textbf{Pass 2}

\textbf{Heuristics}

Guided by the insights gained in Phase 1, you are now free to explore any part of the system. Evaluate the system using each of the following heuristics, which are adapted from Nielsen’s 10 Usability Heuristics[]. For your convenience, a short summary of each heuristic is included below: 

\begin{itemize}
\item Visibility of System Status 

Ensure that users are informed about what the LLM is doing at any given time, such as processing a query or fetching data. 

\item Match Between System and the Real World 

Use language and concepts that are familiar to threat intelligence professionals, avoiding technical jargon that may not be widely understood. 

\item User Control and Freedom 

Provide users with the ability to easily undo or redo actions, and allow them to navigate freely through the system without being locked into specific workflows. 

\item Consistency and Standards 

Ensure that similar actions and terminology are consistent throughout the tool, following industry standards and best practices for threat intelligence platforms. 

\item Error Prevention 

Design the system to prevent errors from occurring in the first place, such as offering confirmation prompts before performing potentially destructive actions. 

\item Recognition Rather Than Recall 

Minimize the user's memory load by making information, actions, and options visible and easily accessible. Provide autocomplete and suggestion features where appropriate. 

\item Flexibility and Efficiency of Use 

Cater to both novice and experienced users by providing shortcuts and advanced options for experienced users while keeping the interface simple for beginners. 

\item Aesthetic and Minimalist Design 

Avoid clutter and provide a clean, focused interface that presents only the necessary information and actions to the user. 

\item Help Users Recognize, Diagnose, and Recover from Errors 

Provide clear and informative error messages, and offer suggestions for how to resolve the issues encountered. 

\item Help and Documentation 

Provide comprehensive help and documentation that is easily accessible. Include tutorials, FAQs, and context-sensitive help within the tool. 
\end{itemize}

\clearpage 
\section*{Appendix}
\subsection*{\textbf{B User Study Briefing}}
Thank you so much for agreeing to participate in this study. As mentioned in the email, we are researchers interested in improving the usability of different LLMs in threat intelligence enrichment. This study will provide valuable information on how to design and develop better support for the use of LLMs in threat intelligence.

Today, I’ll have you use two LLM tools and ask a few questions about your experience using them. During this session, I’ll be recording the computer screen and audio from our conversation. You can let me know at any point if you’d like to stop or pause the recording. 

\subsection*{\textbf{C User Study Task Briefing}}
In this scenario, you have been tasked with enriching threat intelligence events from the provided dataset. These events have to been analyzed by two different LLM tools. I’ll have you use the first tool for about 25 minutes, then I’ll ask you a few questions. Then, I’ll have you use the second tool for about 25 minutes, and I’ll ask you a few more questions.

As you are working with the tools, try to think aloud. Say any questions or thoughts that cross your mind regardless of how relevant you think they are. If you are silent for longer than 30 seconds or so, I’ll gently remind you to KEEP TALKING.

(If participants are silent for more than 30 seconds, raise a "KEEP TALKING" sign)

Prioritized list of tasks:

1. Upload the XML file containing CTI data into the LLM 

2. Direct the LLM to extract all threat intelligence references from the data and categorize them 

3. Identify Key Attacker Activities and Patterns and summarize into a report using the LLM, with the help of the categorized data. Please ensure that the report summarizes all key attacker activities from the event data. 

4. Assess the Quality of the Enrichment. 
Do you have any questions before we begin?

(TURN ON SCREEN AND AUDIO RECORDER!)

\subsection*{\textbf{D Post-Study Questions}}
1.  Which issues did you encounter when using the tool?

2. Which functionalities of the tool did you like/dislike the most?

3. Which functionalities of the tool did you find useful?

4. Were there moments when you were confused?

5. Would you use this tool in your threat intelligence work?

6. How would you compare these tools to other LLMs you have used?

\clearpage 
\section*{Appendix}
\subsection*{\textbf{E Summary of Participants' Experience}}
\begin{table}[ht]
\renewcommand{\arraystretch}{1.3}
\caption{Summary of Participant Experience}
\centering
\begin{tabular}{c|c|c|c|c|c|c}
\hline
\bfseries ID & \bfseries LLM 1 & \bfseries LLM 2 & \bfseries Threat Intelligence Familiarity & \bfseries Automation Tools Familiarity & \bfseries LLM Familiarity & \bfseries Professional Experience(Years)\\
\hline
P01 & ChatGPT & Gemini & 4 & 3 & 2 & 5 \\
P02 & Gemini & Cohere & 3 & 2 & 1 & 3 \\
P03 & Cohere & Copilot & 5 & 4 & 3 & 7 \\
P04 & Copilot & Meta AI & 4 & 3 & 2 & 6 \\
P05 & Meta AI & ChatGPT & 3 & 2 & 1 & 4 \\
P06 & ChatGPT & Gemini & 5 & 4 & 3 & 8 \\
P07 & Gemini & Cohere & 4 & 3 & 2 & 2 \\
P08 & Cohere & Copilot & 3 & 2 & 1 & 3 \\
P09 & Copilot & Meta AI & 5 & 4 & 3 & 5 \\
P10 & Meta AI & ChatGPT & 4 & 3 & 2 & 6 \\
\hline
\end{tabular}
\label{participantsummary}
\end{table}

\end{document}